\documentstyle[11pt,paspconf,psfig]{article}

\def\ha{H\,$\alpha$}

\def\oiii{[O~{\sc iii}]}
\def\oii{[O~{\sc ii}]}

\def\gl{$\lambda$}

\def\kms{km\,s$^{-1}$}

\begin{document}

\title{Tunable filter imaging: structure forming around
a quasar at $z=0.9$} 

\author{Joanne C. Baker}
\affil{MRAO, Cavendish Laboratory, Madingley Road, Cambridge CB3 OHE, UK}

\begin{abstract}

Preliminary results are described for emission-line imaging of the 
field of a quasar at $z=0.9$ with a unique new instrument --- the 
TAURUS Tunable Filter (TTF). At least fourteen \oii-emitting galaxy 
candidates are found within $\pm 750$\kms\ of the quasar redshift with a
significance of $>3\sigma$. 
Another eight candidates ($>3\sigma$) are also found with 
relative velocities $\pm 750-1500$\kms\ from the quasar, 
indicative of a large velocity dispersion in the field. 
Together with the existence of a group of red galaxies in 
the field identified through broad-band imaging, we suggest 
that the TTF data are probing cluster merging which is ongoing  at $z =0.9$. 

\end{abstract}


\section{Introduction}

The search for emission-line galaxies at high-redshift is important for
tracing the history of star formation in the universe.  Variations in the
rate at which galaxies are forming stars are apparent in regions of
different galactic density. For example, lower star formation rates have
been measured in rich galaxy clusters (Balogh et al. 1997) 
than in the field (Cowie et al. 1997). 

As part of a major program (in
collaboration with Bremer, Hunstead and Bland-Hawthorn) to find 
high-redshift galaxies and clusters, deep narrow-band imaging of 
the fields of $z \sim 1$ radio-loud quasars has been obtained 
using a new instrument --- the TAURUS Tunable Filter (TTF). 
At lower redshifts ($z<0.7$) it has been shown that radio-loud 
quasars tend to favour rich environments (Yee \& Green 1987), 
and this makes them good pointers to overdense regions at high redshift.

The TTF is a Fabry-Perot-based imaging system developed recently at 
the  Anglo-Australian Telescope (AAT) enabling, for the first time, 
sensitive narrow-band imaging with a minimum tunable bandpass of 
$\sim 10$\AA\ (Bland-Hawthorn \& Jones 1997). The high throughput 
and narrow bands achievable 
with TTF enable the detection of galaxies at $z \sim 1$ with star 
formation rates of one solar mass per year and higher via their 
\oii\,\gl3727 emission lines. This contribution describes the 
detection of \oii-emitting galaxies with TTF around a quasar at 
$z=0.9$ (our first target).

\section{TTF imaging of a quasar at $z=0.9$}
 
The quasar MRC\,0450$-$221 was observed with TTF at the AAT in February
1997 in a sequence of seven 10\AA-wide bands centred on \oii\ at the
quasar redshift, $z=0.898$. Exposure times were 1000s per band, and the
seeing typically $1.3''$ FWHM. Figure 1 shows sub-images drawn from the 
sequence of TTF images of the $10' \times 10'$ field centred on  
MRC\,0450$-$221. Extended line emission around the quasar (Q), 
reaching up to 1000\,\kms\ redward of the 
nuclear emission, and two emission-line galaxies (G1 \& G2) lying 
within 500\,\kms\ of the quasar redshift are displayed in three panels. 
Each sub-image spans about 500~\kms\ at this redshift.

\begin{figure}[t]
\centerline{\psfig{file=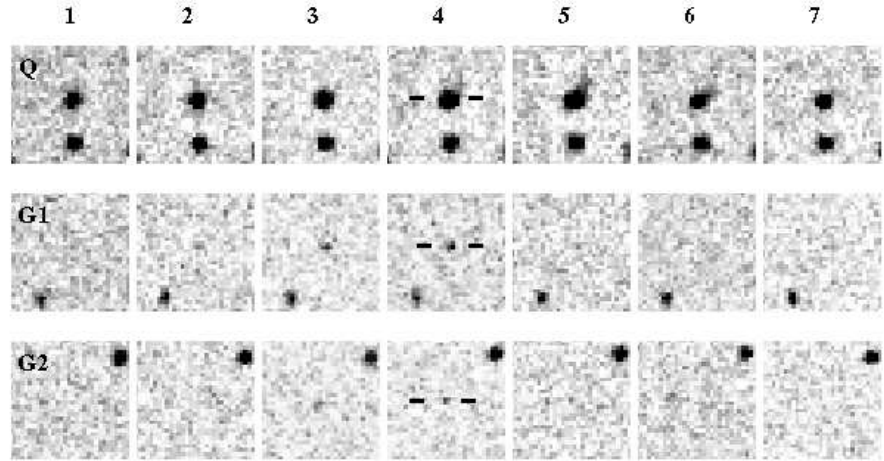,height=6.5cm}}
\caption[]{\footnotesize
Sub-images drawn from a sequence of seven 1000s TTF images of the
field of the $z=0.898$ quasar MRC\,0450$-$221, taken in $10$\AA\ FWHM
adjoining passbands spanning redshifted \oii\ (column 4).  Extended \oii\
emission is clearly seen around the
quasar (Q) elongated in the direction of the radio axis. 
Two galaxies (G1 \& G2) with line emission within 500~\kms\ of
the quasar redshift are also shown.
Each sub-image spans about 500~\kms\ at this redshift. 
}
\label{ttf}
\end{figure} 

\begin{figure}[t]
\centerline{\psfig{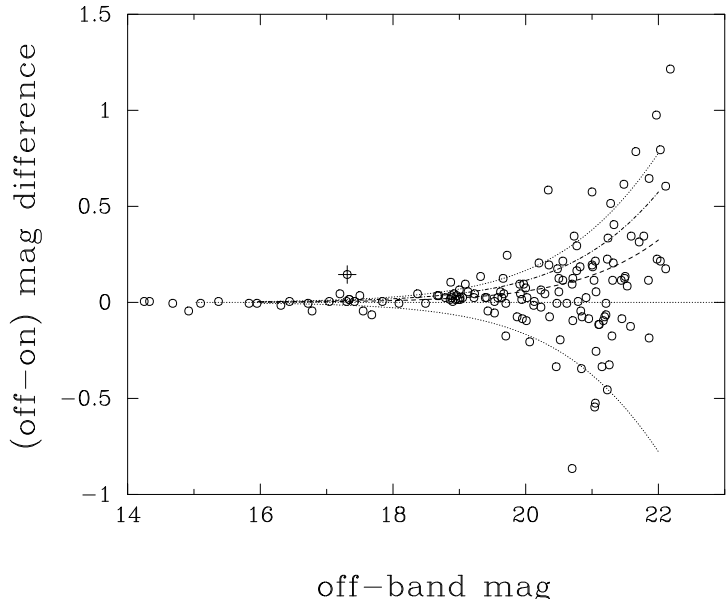} }
\caption[]{\footnotesize
Colour magnitude plot showing magnitude differences between summed TTF
images centred on ($\pm 750$\kms) and off ($\pm 750-1500$\kms) 
the quasar redshift. Signal-to-noise curves ($1,2,3\sigma$ and $-3\sigma$) 
are overlaid. The quasar is identified by a cross. 
}
\label{mags}
\end{figure}

To identify emission-line galaxies at the quasar redshift, average 
images were made centred on ($\pm 750$ \kms) and off ($\pm 750-1500$ \kms) 
the quasar \oii\ wavelength as a first step. Images at the central 
three wavelengths (i.e. bands 3,4,5 in Figure 1) were added together 
to form the on-band image, and the remaining four images co-added 
to form an average off-band image. In practice, a combined image 
of all seven frames was used for the off-band image, as this was 
deeper and the measured magnitudes did not differ significantly 
from the 4-frame off-band image. All the following analysis was 
performed on both off-band frames to confirm the results. 

For each co-added image, aperture photometry was carried out on the
central $7' \times 7'$ using FOCAS. A three-pixel aperture radius 
($2''$) was used, and the
final magnitudes corrected by a constant factor to account for missing
flux. This factor was measured from stars in the field to be a constant 
0.4 mag difference over the whole magnitude range, giving total 
corrected magnitudes which were in excellent agreement with those
measured using a larger aperture. The small aperture size was 
chosen to optimise signal-to-noise for our mainly small and faint targets. 
A finding list was then made from the deep combined image, and the list
was matched to the other catalogues of objects detected by the FOCAS 
algorithms. Photometric uncertainties in the on-band image are 
about 0.2mag at 21 mag ($1\sigma$ at 7070\AA).

Emission-line galaxies have been identified on the basis of the above 
photometry (a range of other photometric techniques were tried, and 
give consistent results).  
Fourteen candidates fainter than 18.5 mag with excess emission $\pm
750$ \kms\ from the quasar redshift (i.e. brighter in the on-band image)
with a significance of $>3\sigma$ are found and are shown on 
the colour-magnitude diagram in Figure 2. These fourteen candidates comprise
20\% of all the objects fainter than 18.5\,mag with positive magnitude 
differences in Figure 2, which clearly exceed the proportion expected 
due to noise alone.

Line emitting galaxies over a large range of velocities relative to the
quasar redshift ($\pm750-1500$ \kms) are also seen in the field, as 
indicated by objects with significant but negative magnitude differences in 
Figure \ref{mags}, i.e. eight objects $>3\sigma$. The identification of such
objects as emission-line galaxies can be confirmed by inspecting 
magnitudes in the seven TTF wavelength bins, such that they appear 
much brighter in one or two bands. Therefore, a large velocity 
dispersion is inferred for the emission-line galaxies in this field, at 
least $\pm 1500$ \kms.

\section{Discussion: subcluster merging at $z=0.9$?}

The number of emission-line galaxies found in our TTF survey ($\sim 22$) 
exceeds the expected value of 1-2 line-emitting field galaxies at $z=0.9$ 
in the volume surveyed, based on the space densities of Cowie et al.
(1997). The detections have been confirmed by visual inspection on 
TTF images and $R$ and $I$ broad-band images of the field. The faint 
magnitudes and broad-band colours of the candidates make it highly 
likely that they are at the quasar redshift rather than being 
intervening \ha\ or \oiii\ emitters (spectroscopic confirmation 
is being sought), although about 3-4 low-redshift contaminants are 
expected in the TTF data according to the numbers of Cowie et al. (1997). 
Assuming the line-emitting objects lie at $z=0.9$, the implied \oii\ 
equivalent widths range between  30-300\AA\ and reach star formation 
rates of a few solar masses per year, consistent with values derived 
for field galaxies (Cowie et al. 1997).

The overdensity of star-forming galaxies in the field of MRC\,0450$-$221 
suggests the existence of a cluster, which is supported by observations of 
a group of red galaxies which is visible to the southeast of and around 
the quasar. The red galaxies ($R-I>1$) have colours consistent with being
passively-evolving ellipticals at $z=0.9$.
An investigation of the relationship between the emission-line galaxies 
identified above and the red galaxy group in the field is ongoing --- 
the candidate \oii-emitters appear to cluster around the quasar,
but more weakly than the red galaxies. It is possisble that
the star-forming galaxies are infalling towards an older cluster core
traced by the red galaxies. Such a picture would be expected at $z \sim 0.9$ 
if clusters form hierarchically.

\section{Conclusions}

Using TTF, twenty-two emission-line galaxies are detected in the 
field of the $z \sim 0.9$ quasar MRC\,0450$-$221 with a significance of
greater than $3\sigma$. The presence of star forming galaxies around the
quasar, together with an overdensity of red galaxies in the field
suggests that the quasar lies in, or near a cluster at $z \sim 0.9$. 
The large velocity dispersion in the field suggests that the line-emitting  
galaxies probed by TTF are infalling onto the older cluster core, and 
perhaps lie in merging subclumps. 

A full analysis of this field will be published shortly (Baker et al.
in prep). TTF observations of more $z\sim1$ quasar fields are 
ongoing and will address the properties of high-redshift clusters 
and their relationship with quasar activity.

\acknowledgments
JCB thanks Joss Bland-Hawthorn and Heath Jones for pioneering
TTF, and the staff of the AAT for technical support.

\end{document}